\def\kms{\relax \ifmmode {\,\rm km\,s}^{-1}\else \,km\,s$^{-1}$\fi}

\def\farcs{\hbox{$.\!\!^{\prime\prime}$}}
\def\arcdeg{\hbox{$^\circ$}}

\def\arcsec{\hbox{$^{\prime\prime}$}}
\def\secd#1.#2{ #1\farcs#2 }               

\def\mincir{\ \raise-2.truept\hbox{\rlap{\hbox{$\sim$}}\raise5.truept
    \hbox{$<$}\ }}
\def\magcir{\ \raise-2.truept\hbox{\rlap{\hbox{$\sim$}}\raise5.truept
    \hbox{$>$}\ }}

\def\nii{[N~{\sc ii}]}
\def\oiii{[O~{\sc iii}]}

\def\oii{[O~{\sc ii}]}
\def\ha{H$\alpha$}
\def\hb{H$\beta$}
\def\hg{H$\gamma$}
 
\def\ariii{Ar~{\sc iii}}

\def\hii{H~{\sc ii}}
\def\heii{He~{\sc ii}}

\def\hei{He~{\sc i}}
\def\sii{[S~{\sc ii}]}
\def\siii{[S~{\sc iii}]}
\def\cbeta{c({H$\beta$})}

\def\nii{[N~{\sc ii}]}

\def\oiii{[O~{\sc iii}]}

\def\oii{[O~{\sc ii}]}
\def\ha{H$\alpha$}
\def\hb{H$\beta$}
\def\hg{H$\gamma$}
\def\hd{H$\delta$}
 
\def\ariii{[Ar~{\sc iii]}}

\def\hii{H~{\sc ii}}
 \def\heii{He~{\sc ii}}

\def\hei{He~{\sc i}}
\def\sii{[S~{\sc ii}]}
\def\siii{[S~{\sc iii}]}

\documentclass{aa}
\usepackage{graphicx}
\usepackage{txfonts}
\usepackage{longtable}
\begin{document}

\title{The metallicity gradient of M~33:
\\ chemical abundances of \hii\ regions \thanks{Based on
observations obtained at the 4.2m~WHT telescope operated on the island
of La Palma by the Isaac Newton Group in the Spanish Observatorio del
Roque de Los Muchachos of the Instituto de Astrofisica de Canarias.}}
\titlerunning{The gradient of M33}
\author{
L. Magrini\inst{1},
J. M. V\'{\i}lchez \inst{2}.
A.   Mampaso   \inst{3},
R.L.M. Corradi \inst{4,3},
P. Leisy\inst{4,3}
}
\authorrunning{Magrini et al.}

   \offprints{L. Magrini\\
e-mail: laura@arcetri.astro.it}

\institute{INAF--Osservatorio Astrofisico di Arcetri,
              Largo E. Fermi, 5, 50125 Firenze, Italy
\and
Instituto de Astrof\'{\i}sica de Andaluc\'{\i}a  (CSIC)
Apartado de Correos 3004, 18080 Granada,  Spain
\and
Instituto de Astrof\'{\i}sica de Canarias, c. V\'{\i}a L\'actea s/n,
38200, La Laguna, Tenerife, Canarias, Spain
\and
Isaac Newton Group of Telescopes, Apartado de Correos 321, 38700 Santa
Cruz de La Palma, Canarias, Spain
}

\date{Received ; accepted}


\abstract{
We present spectroscopic observations of a sample of 72 emission-line
objects, including mainly \hii\ regions, in the spiral galaxy M~33.
Spectra were obtained with the multi-object, wide field spectrograph AF2/WYFFOS at
the 4.2m~WHT telescope.  Line intensities, extinction, and electron
density were determined for the whole sample of objects.}
{The aim of the present work was to derive chemical and physical
parameters of a set of \hii\ regions, and from them the metallicity
gradient.}
{Electron temperatures and chemical abundances were derived for the
14 \hii\ regions where both \oii\ and \oiii\ emission line fluxes were
measured, including the electron temperature sensitive emission line
\oiii\ 436.3~nm and in a few cases \nii\ 575.5~nm.  The ionization
correction factor (ICF) method was used to derive the total chemical
abundances.}
{The presence of abundance gradients was inferred from the
radial behaviour of several emission-line ratios, and accurately
measured from chemical abundances directly derived in 14 \hii\
regions. The oxygen abundances of our \hii\ regions, located in the
radial region from $\sim$2 to $\sim$7.2~kpc, gave an oxygen gradient
-0.054$\pm$0.011~dex~kpc$^{-1}$ }
{The overall oxygen gradient for M~33 obtained using ours and previous
oxygen determinations in a large number of \hii\ regions with direct
electron temperature determination as well as abundance in young stars presented a two
slope shape: -0.19~dex~kpc$^{-1}$ for the central regions (R$<$3~kpc),
and -0.038 ~dex~kpc$^{-1}$ for the outer regions (R$\geq$3~kpc).  }


\maketitle
%

\section{Introduction}
\label{sect_intro}

The galaxy M~33 (NGC~598) is the third-brightest member of the Local
Group.  Its closeness (840~kpc, Freedman et al.~\cite{freedman91}),
its large angular size (optical size 53'$\times$83', Holmberg
\cite{holmberg58}), and its intermediate inclination ($i$=53\arcdeg)
make it particularly suitable for studies of spiral structure,
interstellar medium (ISM), and of stellar populations (van den Bergh
\cite{vdb00}).
Being a late-type spiral galaxy, it has a rich population of \hii\
regions.
A catalog of large-sized \hii\ regions was given by
Courtes et al.~(\cite{courtes87}), whereas the positions of
compact \hii\ regions were published by Calzetti et
al.~(\cite{calzetti95}).
One of the most recent catalogs, published
in two part by Wyder et al.~(\cite{wyder97}) and Hodge et
al.~(\cite{hodge99}), lists a large number of new \hii\ regions.  Observations with the
2.5~m~INT telescope at La Palma, Spain, covering an area of
approximately 0.6 square degrees,  produced the most complete
catalog of \hii\ regions in this galaxy
(Cardwell et al.~\cite{cardwell00}).  Recently, the
Local Group Census consortium (LGC {\tt http://www.ing.iac.es/$\sim$rcorradi/LGC/},
cf. Corradi \& Magrini~\cite{corradi06}) obtained new
imaging observations of M~33.  These new deep data cover about 2
square degrees and are allowing to discover a conspicuous number of new
\hii\ regions at large galactocentric distances.

The existence of chemical abundance gradients in M~33 was known from
long time, but its origin, shape and magnitude are still open issues.
Aller (\cite{aller42}) noted that among \hii\ regions in M33, those
far from the center had much larger \oiii/\oii\ emission line ratios
than those near the center.  Searle (\cite{searle71}) presented
spectrophotometry of eight \hii\ regions in M~33 and recognized that a
radial abundance gradient could explain this trend of flux ratios.

Further spectroscopic studies of \hii\ regions were carried on by
Smith~(\cite{smith75}), Kwitter \& Aller~(\cite{kwitter81}), and by
V\'{\i}lchez et al.~(\cite{vilchez88}). Garnett et al.~(\cite{garnett97})
in a recompilation of previous observations, obtained an overall O/H
gradient of -0.11$\pm$0.02~dex~kpc$^{-1}$.  Willner \&
Nelson-Patel~(\cite{willner02}) derived neon abundances of 25 \hii\
regions from infrared lines, obtaining a Ne/H gradient
-0.05$\pm$0.02~dex~kpc$^{-1}$.
A new recompilation of young stars and \hii\ regions abundance determinations
was done also by P{\'e}rez-Montero \& Di{\'a}z~(\cite{perez05}).
Recently, Crockett et
al.~(\cite{crockett06}) derived even shallower gradients (Ne/H
-0.016$\pm$ 0.017~dex~kpc$^{-1}$, and O/H
-0.012$\pm$0.011~dex~kpc$^{-1}$) using chemical abundances of 6 \hii\
regions with derived electron temperature.

The abundance gradient of M~33 was also investigated using planetary
nebulae (PNe), young giant stars, and red giant branch (RGB) stars.
PNe chemical abundances were studied by Magrini et al.~(\cite{m04})
via optical spectroscopy and photoionization modelling and by
Stasinska et al.~(\cite{stasinska06}) with direct electron temperature
measurements of PNe located in the inner regions of M~33.  Stellar
abundances were obtained by Herrero et al.~(\cite{herrero94}) for
AB-supergiants, McCarthy et al.~(\cite{mccarthy95}) and Venn et
al.~(\cite{venn98}) for A-type supergiant stars, and Monteverde et
al. (\cite{monteverde97}, \cite{monteverde00}) and Urbaneja et
al.~(\cite{urbaneja05}) for B-type supergiant stars.  The O/H gradient
as derived by Urbaneja et al.~(\cite{urbaneja05}) is
-0.06$\pm$0.02~dex~kpc$^{-1}$.  The metallicity measured in Cepheids
identified in the inner regions of M~33 by Beaulieu et
al.~(\cite{beaulieu06}) suggested an O/H gradient of
$-0.16$~dex~kpc$^{-1}$.  Metallicities of old stellar populations were
derived via deep CCD photometry and colour magnitude diagrams by
Stephens \& Frogel (\cite{stephens02}), Kim et al. (\cite{kim02}),
Galletti et al. (\cite{galletti04}), Tiede et al.~(\cite{tiede04}),
Brooks et al. (\cite{brooks04}), and Barker et al.  (\cite{barker06}).
The [Fe/H] gradient computed by Barker et al.  (\cite{barker06})
including all previous iron determinations in RGB stars was
-0.07$\pm$0.01~dex~kpc$^{-1}$.

The aim of the present paper is to investigate the chemical and
physical properties of a large sample of \hii\ regions of different
sizes, galactocentric distances and excitations and to study their
behaviour through the disk.

The paper is organized as follows: in Sect.~\ref{sect_obs} we present
observations and data reductions, while in Sect.~\ref{sect_ext} we
discuss the behaviour of the extinction across the disk.
Emission-line ratio diagrams are shown in Sect.~\ref{sect_dia} and the
physical and chemical properties of the studied \hii\ regions are presented in
Sect.~\ref{sect_chem}.  The metallicity gradients and their
implication in the evolution of M~33 are discussed in
Sect.~\ref{sect_grad} and in Sect.~\ref{sect_discu}.  Finally in
Sect.~\ref{sect_conclu} we summarize our conclusions.


\section{Observations and data reduction}
\label{sect_obs}

\begin{figure*}
\resizebox{\hsize}{!}{\includegraphics[angle=0]{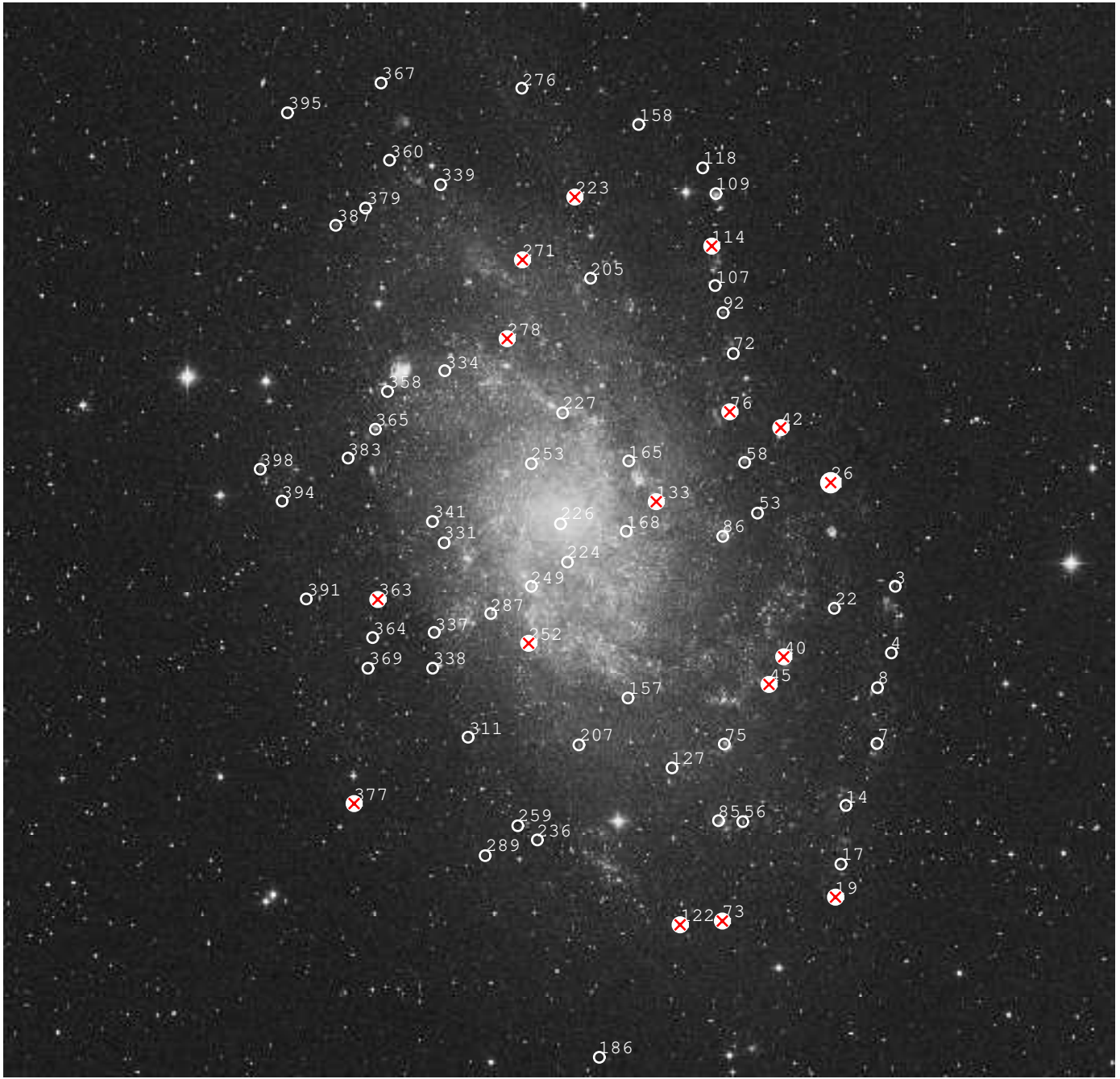}}
\caption{M33 DSS 1\arcdeg $\times$ 1\arcdeg image indicating  the location
of the 72 observed emission-line objects (North is up and East to the left).
The labeled id correspond to the first column of Table~\ref{Tab_pos_hii}. \hii\ regions
with measured chemical abundances  are marked with a cross.}
\label{Fig_m33_hii}
\end{figure*}

Seventy-two emission-line objects (see Fig.~\ref{Fig_m33_hii}) were
observed on November 18th and 19th, 2004 with AF2/WYFFOS, the
multi-object, wide field, fibre spectrograph working at the prime
focus of the 4.2~m William Herschel Telescope (La Palma, Spain)
equipped with an Atmospheric Dispersion Corrector that corrects for
the effect of atmospheric dispersion. We used a 1024$\times$1024 TEK
CCD, and the Small Fibre module which contains 150 science fibres with
1.6\arcsec\, diameter (90 $\mu$m) projected on the sky.  The WYFFOS
spectrograph was used with a single setup: R600B grating (600 line
mm$^{-1}$) providing a dispersion of 3.0 \AA~pix$^{-1}$.  The
observations were repeated with two different incidence angles and
consequently two different central wavelengths were obtained: 513~nm
and 574~nm.  The resulting total spectral coverage ranged from
approximately 370.0~nm to 750.0~nm, and included the basic lines
needed for the classification of the objects and the determination of
their physical and chemical properties.  The response of the
instrument in the bluer part of the spectra ($<$400.0~nm) was
inadequate and thus not suitable for further analysis.  We have
considered the spectral range from 410.0~nm to 750.0~nm for the
subsequent work.

The positions of our targets were obtained
from the INT+WFC images by Magrini et al. (\cite{magrini00}) and from
Local Group Census (LGC) observations (Corradi \& Magrini
\cite{corradi06}).  The selection criteria for the observed sample of
emission-line objects, including mostly \hii\ regions, but also PNe
and SNRs, were the following: {\em i)} uniform distribution over the
face of the galaxy; {\em ii)} large range of sizes and shapes; {\em
iii)} bright \oiii\ $\lambda$~500.7~nm emission in order to make
possible the chemical abundance determination.
We aimed to contribute to the study of the metallicity gradient
in M~33 with a sample of \hii\ regions of different morphological
types and sizes, selecting for observation \hii\ regions without
previous spectroscopic data, avoiding to repeat the study of already
well analyzed giant \hii\ regions.  The coordinates, classifications
and fluxes in previous catalogs, when available, are shown in
Table~\ref{Tab_pos_hii}.  The 13 emission-line objects not identified
in previous works, mostly small isolated \hii\ regions, are labeled
with LGC-HII-n, standing for \hii\ regions discovered by the LGC
project. The \ha+\nii\ fluxes of these \hii\ regions were computed
using aperture photometry in the INT+WFC images by Magrini et
al.~(\cite{magrini00}).

 A total of 6 science exposures of 1800~s each were taken with the
central wavelength $\lambda$ 513~nm, and with a mean airmass 1.1,
and 3 exposures of 1800~s at $\lambda$ 574~nm, with a mean airmass 1.4.
The seeing during the observations was 1.5\arcsec\,
during the first night and $<$1\arcsec\, in the second night.
Several
offset sky exposures using the same fibre configuration were taken
before and after the M~33 observations to perform sky subtraction.
Twilight sky and tungsten lamp exposures were taken as flat fields,
while helium and neon lamp exposures were used for wavelength calibration.

The data were reduced using the {\sc IRAF} \footnote{{\sc IRAF} is
distributed by the National Optical Astronomy Observatories, which are
operated by the Association of Universities for Research in Astronomy,
Inc., under cooperative agreement with the National Science
Foundation.} multi-fibre spectra reduction package {\sc DOFIBER}.  The data
reduction of a similar set of observations, and in particular the sky
subtraction procedure, were described in detail in Magrini et
al. (\cite{magrini03}).  Here we remind briefly the sky subtraction
process.  The sky spectra were taken in two different ways: {\em i)}
with sky fibres taken during the exposures of M~33, that served to
monitor the relative intensity of the atmospheric emission lines,
which are known to vary during the night, {\em ii)} with offset-sky
spectra that allowed to control the variations in the instrumental
profiles of the position of each individual fibre in the focal plane.
The final sky subtraction was done using the offset-sky spectra, after
correcting their relative intensity of the atmospheric lines using the
mean sky spectrum computed from the sky fibres observed at the same
time as the science targets.  Relative flux calibration was
obtained taking spectra with several fibres of the spectrophotometric
standard stars HD~93521 and BD+25~4655 (Oke
\cite{oke90}).
All standard stars were observed with airmass $<$1.1.
The feasibility of the use of a mean sensitivity
function for all the fibres was demonstrated by Magrini et
al. (\cite{magrini03}).  Emission line fluxes were measured using the
{\sc IRAF} SPLOT package. Errors in the fluxes were calculated taking
into account the statistical error in the measurement of the fluxes,
as well as systematic errors of the flux calibrations, background
determination, and sky subtraction.  The observed emission line fluxes
and their errors are listed in Tab.~\ref{Tab_hii_flux}.

\section{The  extinction}
\label{sect_ext}

The observed line fluxes were corrected for the effect of the
internal and interstellar extinction.  The extinction law of Mathis
(\cite{mathis90}) with $R_V$=3.1 was used.  The \cbeta, which measures
the logarithmic difference between the observed and un-reddened \hb\
fluxes, was determined comparing the observed Balmer I(\ha)/I(\hb)
ratio with its theoretical value.
The effect of the underlying absorption
from the continuum below the Balmer lines was neglected.
The extinction
coefficients \cbeta\ for the complete sample of \hii\ regions are
shown in Tab.~\ref{Tab_hii_flux}.

\begin{figure}
\resizebox{\hsize}{!}{\includegraphics[angle=-90]{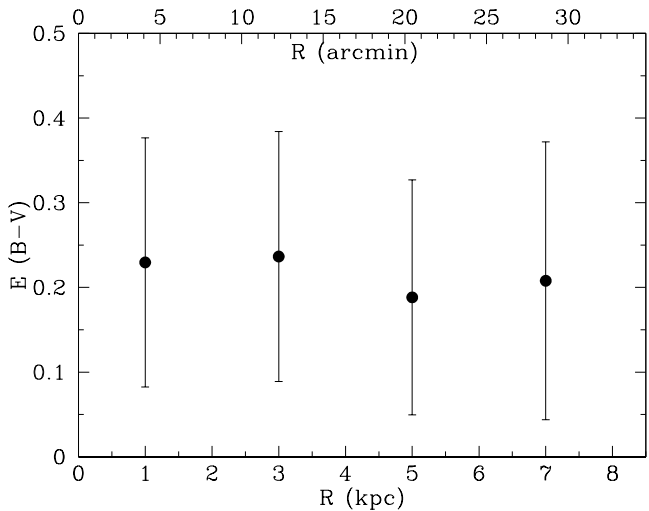}}
\caption{The radial dependence of the extinction, converted form \cbeta\ to  E(B-V) with the
relation \cbeta=0.4 R$_{\beta}$ E(B-V), where R$_{\beta}$=3.7.
The {\em filled circles} are the mean extinction in each 2~kpc wide bin.}
\label{Fig_ext}
\end{figure}

In Fig.~\ref{Fig_ext}, the radial dependence of the extinction,
converted in E(B-V) with the relation \cbeta=0.4 R$_{\beta}$ E(B-V),
where R$_{\beta}$=3.7, is reported. We remind that in this plot and in
the following diagrams the adopted distance to M~33 is 840~kpc
(Freedman et al.~\cite{freedman91}).  The figure shows the histogram
of the mean extinction in each 2~kpc bin from the centre of the
galaxy.  The error is the {\em rms} scatter of the average of the
values of the extinction in each bin.  Note an almost uniform trend of
the extinction, with an average value across the whole disk of
$<$E(B-V)$>$=0.22$\pm$0.15, corresponding to A$_V$=0.65.

\begin{figure}
\resizebox{\hsize}{!}{\includegraphics[angle=-90]{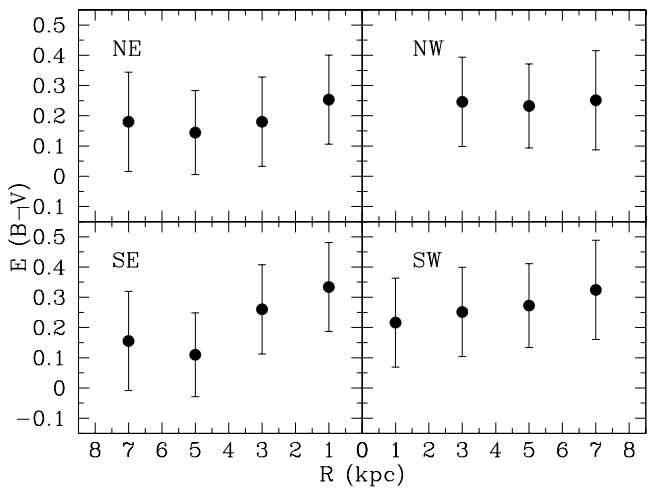}}
\caption{ The radial dependence of the extinction in the four quadrants of M~33: NE north-east, NW north-west,
SE south-east, and SW south-west.  Note that in the NE and SE
quadrants the x axes are inverted to reproduce, together with the
other quadrants, the global radial behaviour of the extinction in the
galaxy, with the inner regions of the galaxy located in the centre of
the plot.  The {\em filled circles} show the mean extinction in each
2~kpc wide bin.}
\label{Fig_ext_quarter}
\end{figure}

In Fig.~\ref{Fig_ext_quarter}, the radial dependence of the extinction
in four quadrants of M~33 is also shown: NE north-east, NW north-west,
SE south-east, and SW south-west.  The extinction was computed as
described for Fig.~\ref{Fig_ext}.  The point in the first bin of the
SW quadrant is not very representative being the average of only two
\hii\ regions extinctions.  Note that the nebulae located in the
eastern part of the galaxy present a lower average extinction.  This
might be attributed to the effect of the inclination of M~33.  The
eastern parts of the galaxy are indeed the farther ones, and because
of an effect of projection, dust filaments are observed below the
\hii\ regions, while in the western regions, because of the same
effect, the dust is located above the \hii\ regions, thus enhancing
the total extinction.

Previous work by Viallefond \& Goss (\cite{viallefond86}) and by
Israel \& Kennicutt (\cite{israel80}) examined the extinction
behaviour in a sample of \hii\ regions of M~33.  They both found that
the extinction derived from Balmer lines is smaller than the
extinction obtained from radio free-free and Balmer line
comparison. This is due to a non uniform distribution of dust inside
the \hii\ regions, together with the different part of the nebula
monitored by the two methods: the first method is more sensitive to
the extinction in the outer zones while the latter in the internal
zones.  They found a radial dependence of the extinction:
\begin{equation}
A_v=2.3 - 0.07 \times R
\end{equation}
where R is the galactocentric distance expressed in arcmin.

Devereux et al.~(\cite{devereux97}) on the other hand, analyzing a
sample of small \hii\ regions, found an average extinction
A$_V$$\sim$1~mag from radio to H$\alpha$ comparison, with no
systematic dependence on radius. They explained their results
contrasting those by Israel \& Kennicutt (\cite{israel80}) as if the
bright \hii\ regions show a radial dependence of extinction while the
fainter ones do not.  Our data are consistent with the results of
Devereux et al.~(\cite{devereux97}).  The absolute value of our
average extinction, A$_V$=0.65, based on Balmer lines ratios and thus
involving only the outer parts of the \hii\ regions, is slightly
lower, as expected, than the value derived by radio and optical Balmer
line comparison, A$_V$$\sim$1.

\section{Emission-line ratio diagrams}
\label{sect_dia}

\begin{figure}
\resizebox{\hsize}{!}{\includegraphics[angle=-90]{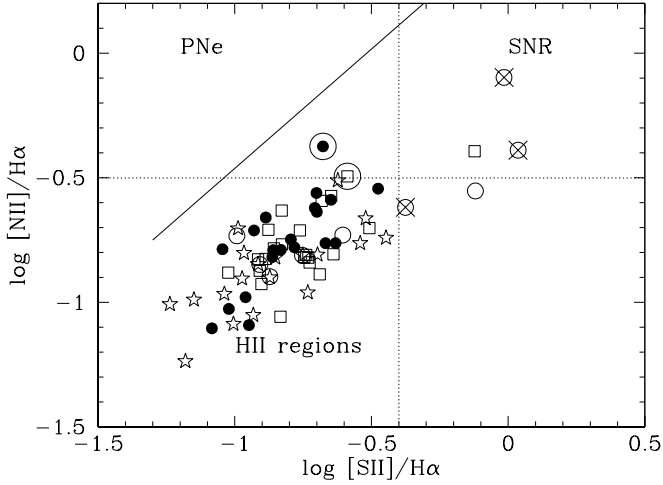}}
\caption{\sii/\ha\ vs. \nii/\ha\ diagnostic diagram.
The different morphological types of emission-line objects are
indicated by the following symbols: small isolated ({\em filled
circles}), knots in extended sources ({\em empty squares}), diffuse
and extended objects ({\em stars}), rings ({\em empty circles}).
The candidate SNRs and PNe known in the literature are marked additionally
with  further crosses and circles, respectively.
Errors on the logarithmic ratios are the order of the size of the
symbols.  The dashed lines delineate the limiting values that best
separate the photoionized and shock-ionized objects (Galarza et
al.~\cite{galarza99}).
Photoionized regions are located in the first quadrant (bottom left).
The continuous line marks the lower limit of
the PN region, as derived for a large sample of Galactic PNe by
Riesgo \& L\'opez~(\cite{riesgo06}).}
\label{Fig_dia}
\end{figure}

In order to minimize the effect of the reddening correction, we have
chosen, when possible, ratios of emission-lines which are close to
each other in wavelength:
\oiii $\lambda$500.7~nm/\hb, \nii $\lambda$658.4~nm/\ha,  and \sii $\lambda\lambda$671.7-673.0~nm/\ha.
The \nii\ and \sii\ emission lines with respect to \ha\ are useful to
separate photoionization and shock-ionization mechanisms, and thus to
distinguish among \hii\ regions, PNe and SNRs.
In Fig.~\ref{Fig_dia} the observed \sii/\ha\, and \nii/\ha\, ratios are
plotted.
The two dashed lines delineate the limiting values that best separate the
photoionized and shock-ionized objects (Galarza et
al.~\cite{galarza99}).  The location of the horizontal dashed line
relative to \nii/\ha\, was shifted to a lower value than indicated
by Galarza et al.~(\cite{galarza99}) for M31 (from $\log$(\nii/\ha)=-0.3 to $\log$(\nii/\ha)=-0.5)
to take into account the lower metallicity of M33.

The vertical line, \sii/\ha, was maintained as in M31, since this flux
ratio is less sensitive to metallicity.  The continuous line marked
the lower limit of the PN regions, as derived for a large sample of
Galactic PNe by Riesgo \& L\'opez~(\cite{riesgo06}).  The
emission-line objects were plotted with different symbols according to
their morphology: small isolated regions with high surface brightness
({\em filled circles}), high surface brightness knots within extended
sources ({\em empty squares}), diffuse and extended objects ({\em
stars}), ring-like nebulae ({\em empty circles}).

From Fig.~\ref{Fig_dia}, we noticed that most of the observed
emission-line objects are located in the \hii\ regions area.  The
correlation of the ratios \sii/\ha\, and \nii/\ha\, of \hii\ regions
appearing in Fig.~\ref{Fig_dia} is independent of morphological types
and depends on the metallicity and excitation of the nebulae, as
demonstrated by the photoionization modelling by Viironen et
al.~(\cite{viironen07}).  Five of the emission-line objects have
ratios consistent with being SNRs.  Among them, four have ring-like
morphologies, and three of them were indeed already classified as SNRs
(see Table~\ref{Tab_pos_hii}).  Among the candidate PNe present in the
literature, one of them, PN CDL04~52, appeared from our LGC images to
be a knot of an extended \hii\ region, whereas the position of PN
MCM00~66, in the \sii/\ha\, vs. \nii/\ha\, diagram is well detached
from the area of \hii\ regions and it can be considered a true PN.  We
excluded then from the following analysis the three known SNRs
(M33SNR~15, 25, 64), as well as the two new candidate SNRs LGC-HII~5,
BCLMP207a, and the PN MCM00~66.

\begin{figure}
\resizebox{\hsize}{!}{\includegraphics[angle=-90]{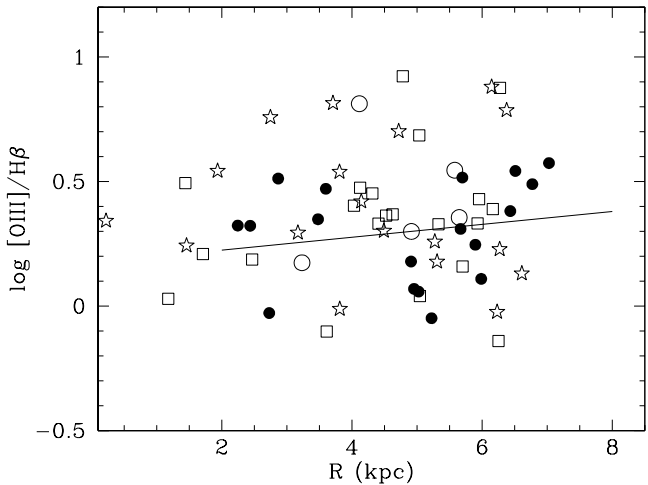}}
\caption{Radial distribution of \oiii/\hb\ line ratio in different morphological
types of \hii\ regions. Symbols and errors as in Fig.~\ref{Fig_dia}.
The solid line is the fit for the isolated  high surface brightness nebulae only
({\em filled circles}). }
\label{Fig_oiii}
\end{figure}

The radial emission-line ratio distributions presented in
Figs.~\ref{Fig_oiii}, \ref{Fig_nii}, and \ref{Fig_sii} show a large
scatter at each radius, even if some radial trends can be identified.
The scatter cannot be attributed to errors in the line fluxes (smaller
than the symbols as plotted in Figs.~\ref{Fig_oiii} to ~\ref{Fig_sii})
nor to errors on the position of the nebulae within the galaxy.  In
fact, the well known inclination (53\arcdeg$\pm$1\arcdeg) of M33
allowed an accurate determination of galactocentric distances. Typical
errors on the deprojected galactocentric distances associated with the
uncertainty on the inclination were less than 0.1~kpc.

Thus the scatter must be real and represents the fact that each
emission-line ratio depends on various characteristics of the nebula,
which are not necessarily function of the galactocentric distance
(cf. Blair \& Long \cite{blair97}). In particular, it can be ascribed
to internal variations in temperature, excitation, and density of the
nebula, or to the nature of the ionizing sources, as the relative
weight of the O star population which might affect significantly the
resulting nebular spectrum.

Fig.~\ref{Fig_oiii} shows the radial variation of \oiii/\hb.  This
ratio measures the excitation of the radiation field within the
ionized nebula or, equivalently, the temperature of the central
stars. It is also sensitive to chemical composition (Searle
\cite{searle71}).  An increase in \oiii/\hb\ towards the external
regions is known to exist in disk galaxies and it is considered to be
an evidence for a negative oxygen gradient (Scowen et
al.~{\cite{scowen92}, Zaritsky et al.~\cite{zaritsky90}). However, an
empirical correlation between metallicity and the observed \oiii/\hb\
line ratio cannot be adopted since this line ratio measures only one
ionization state of oxygen, O$^{2+}$, and in low-excitation objects,
such as \hii\ regions, this might represent a small contribution to
the total oxygen abundance.  For M33, the expected increase in
excitation with galactocentric distance is evident only in small
isolated \hii\ regions.  The diffuse nebulae, including their compact
portions, as well as the ring-like nebulae, reproduce only marginally
this trend and appear to introduce more scatter in the range covered
by the high surface brightness sources, as also observed in M31 \hii\
regions (Galarza et al.~\cite{galarza99}).  The solid line shown in
Fig.~\ref{Fig_oiii} represents the best-fit to the isolated high
surface brightness sources only ({\em filled circles}).

\begin{figure}
\resizebox{\hsize}{!}{\includegraphics[angle=-90]{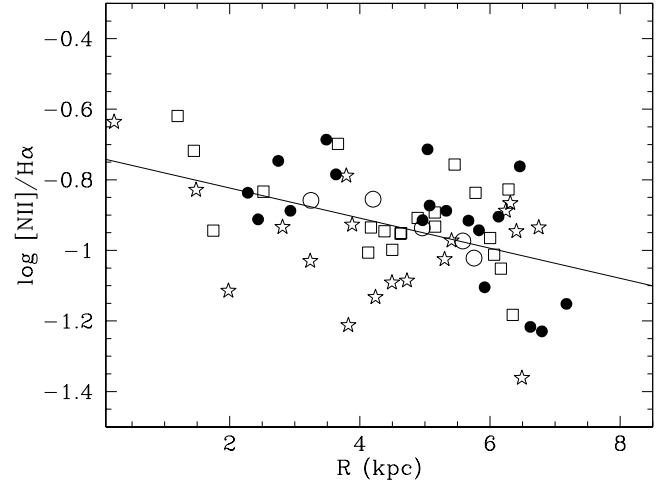}}
\caption{Radial distribution of \nii/\ha\ line ratio.
Symbols and errors as in Fig.~\ref{Fig_dia}.}
\label{Fig_nii}
\end{figure}

The radial plot of \nii\ 658.4 relative to \ha\, is shown in
Fig.~\ref{Fig_nii}.  A decreasing trend toward the outer disk of M33
is evident, mainly due to the abundance gradient. In fact, the
intensity of \nii\ lines is not affected by collisional de-excitation,
consequently is insensitive to electron density, while it is mostly
dependent to the nitrogen abundance (cf. Viironen et
al. \cite{viironen07}).  Note however that the scatter in the relation
is huge, approximately a factor 3 at given radius, rendering this line
ratio useful only for a qualitatively detection of the abundance
effect on the gradient.

\begin{figure}
\resizebox{\hsize}{!}{\includegraphics[angle=-90]{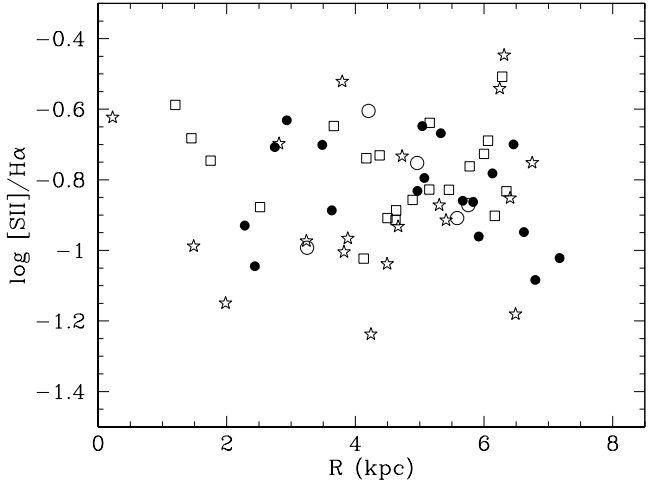}}
\caption{Radial distribution of \sii/\ha\ line ratio.
Symbols and errors as in Fig.~\ref{Fig_oiii}.}
\label{Fig_sii}
\end{figure}

The radial distribution of \sii\ 671.7, 673.1 relative to \ha\, is
shown in Fig.~\ref{Fig_sii}.  There is no clear radial trend, as
expected by theoretical modeling by e.g. Domgorgen \& Mathis
(\cite{domgorgen94}).

\section {Physical and chemical properties}
\label{sect_chem}
\subsection{Electron density and temperature}

The electron densities, n$_e$, were measured in almost all the
analyzed emission-line objects using the \sii\ I(671.7)/I(673.1)
ratio, which is sensitive to n$_e$ because of the collisional
de-excitation of the 671.7~nm transition, but almost insensitive to
the electron temperature.  We used a five-level atom statistics to
compute n$_e$, assuming the electron temperature as measured from the
(I(500.7)+ I(495.9))/I(436.3) ratio, when available, and T$_e$=10000~K
in the other cases.  For four \hii\ regions we gave a direct
measurement of the electron density and for the remaining ones we gave
an upper limit.

Most of our \hii\ regions have electron densities or their upper
limits consistent with the low density limit for nebulae, i.e. n$_e <$
100 cm$^{-3}$, including the PN candidate (MCM00~66) and the five
supernova remnants (SNR M33SNR~15, 25, 64, LGC-HII-5, BCLMP207a)
observed in our sample.  Only three \hii\ regions have a higher n$_e$,
namely LGC-HII-1, BCLMP~220, and BCLMP~705.  Electron density values
or their upper limits are listed in Tab.~\ref{Tab_hii_flux}.

Electron temperatures, T$e$, were derived using the ratio \oiii\
(I(500.7)+ I(495.9))/I(436.3) in 15 \hii\ regions.  In three of them,
where \nii\ 575.5~nm was detectable, we also derived T$e$ using the
ratio \nii\ (I(658.4)+ I(654.8))/I(575.5), finding agreement, within
the errors, between T$_{e{\rm [N~{\rm II}]}}$ and T$_{e{\rm [O~{\rm
III}]}}$.  In addition, an upper limit to T$_e$ was given in other 7
\hii\ regions (see Table~\ref{Tab_hii_flux}).

In Fig.~\ref{Fig_temp}, the derived T$_{e{\rm [O~{\rm III}]}}$ ({\em
filled circles}) and T$_{e{\rm [N~{\rm II}]}}$ ({\em empty circles})
are plotted versus the deprojected galactocentric distance of the
\hii\ regions.  The plot shows the presence of a gradient of
electron temperatures across the galaxy, with a slope of
570$\pm$130~K~kpc$^{-1}$.  This is clearly related to the existence of
an abundance gradient, since metal poorer \hii\ regions have a higher
electron temperature due to the less effective cooling.

\begin{figure}
\resizebox{\hsize}{!}{\includegraphics[angle=-90]{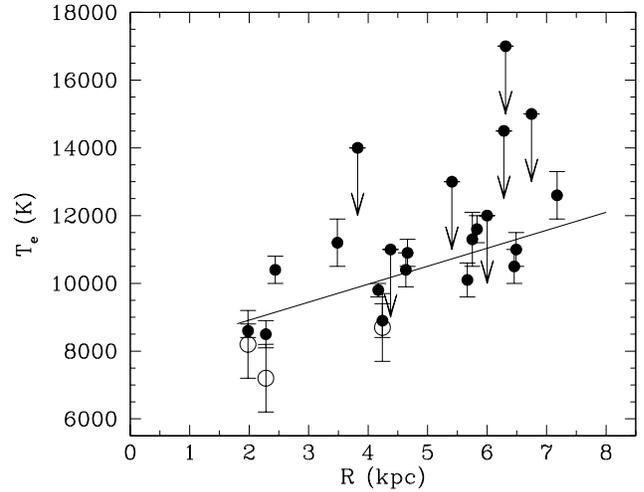}}
\caption{Electron temperature, T$_e$, as a function of galactocentric distance:
T$_{e{\rm [O~{\rm III}]}}$ ({\em filled circles}) and T$_{e{\rm [N~{\rm II}]}}$ ({\em
empty circles}). Upper limits to the T$_{e{\rm [O~{\rm III}]}}$ are marked with an arrow.
The solid line is the weighted linear least-squares
fit to the T$_{e{\rm [O~{\rm III}]}}$ values (excluding the upper limits).}
\label{Fig_temp}
\end{figure}

For the \hii\ regions where only T$_e$\oiii\ was measured we adopted
the relationships by Izotov et al.~(\cite{izotov05}) for high Z
galaxies, 12 + $\log$(O/H)$>$8.2, following their definition:
\begin{equation}
 T_{e {\rm [O~{\rm II}]}} = 2.967 + T_{e {\rm [O~III]}} \times (-4.797 + 2.827 T_{e {\rm [O~III]}})
\end{equation}
and
\begin{equation}
 T_{e  {\rm [S~III]}} = 1.653 + T_{e {\rm [O~III]}} \times (-2.261 + 1.605 T_{e {\rm [O~III]}})
\end{equation}

where T$_{e{\rm [O~{\rm II}]}}$, T$_{e{\rm [O~{\rm III}]}}$, and
T$_{e{\rm [S~{\rm III}]}}$ are both expressed in 10$^4$ K unit.
Following Izotov et al.~(\cite{izotov05}), we then adopted T$_{e{\rm
[O~{\rm II}]}}$ for the calculation of N$^+$, O$^+$, S$^+$ abundances
and T$_{e{\rm [S~{\rm III}]}}$ for the calculation S$^{2+}$ and
Ar$^{2+}$ abundances, while T$_{e{\rm [O~{\rm III}]}}$ directly
measured was used for O$^{2+}$ and He${^+}$ abundances.

\subsection{Chemical abundances}

The spectral coverage of our observations allows the determination of
ionic and total abundances of several elements, namely He/H, N/H, O/H,
S/H, and Ar/H. In order to limit the uncertainty due to the use of
ionization correction factors (ICFs), we considered only the spectra
where both \oii\ and \oiii\ (including the temperature sensitive line,
\oiii\ 436.3~nm) emission lines are measured with sufficient accuracy
(S/N$>$3).  In fact, \hii\ regions are generally low-excitation
objects and a relevant percentage of their oxygen is only once
ionized. Consequently \oii\ emission lines are very relevant in the
calculation of the total oxygen abundance.  In addition, the ICFs used
to compute the total abundance of the other chemical elements are all
expressed as a function of both oxygen ionization stages, either
O$^{+}$/(O$^{+}$+O$^{2+}$) or O$^{2+}$/(O$^{+}$+O$^{2+}$).  We used
\hei\ $\lambda$ 587.6~nm and \heii\ $\lambda$ 468.6~nm for the
determination of the helium abundance, \nii\ $\lambda
\lambda$ 654.8, 658.4~nm for the nitrogen abundance, \oii\ $\lambda
\lambda$ 732.0, 733.0~nm and \oiii\ $\lambda \lambda$ 495.9, 500.7~nm
for the oxygen abundance, \sii\ $\lambda \lambda$ 671.7, 673.1~nm and
\siii\ $\lambda$ 631.2~nm for the sulfur abundance, and
\ariii\ $\lambda$ 713.5~nm for the argon abundance.

Ionic O$+$ abundances are usually computed from the
\oii\ $\lambda\lambda$ 372.7, 372.9~nm line fluxes.
However, the \oii\ $\lambda\lambda$ 732.0, 733.0~nm lines should be
used when, as in our case, the brighter emission lines
$\lambda\lambda$ 372.7, 372.9~nm are not available (c.f. Izotov et
al.~\cite{izotov05}). A comparison of O$+$ abundances derived from
both sets of lines in a sample of about 200 SDSS \hii\ galaxies is
presented by Kniazev et al.~(\cite{kniazev04}). They found a weighted
mean of the differences between the two sets of abundances
$\log$(O/H)$_{372.7,372.9}$ -$\log$(O/H)$_{732.0,
733.0}$=0.002$\pm$0.002 dex, with a rms scatter of 0.02 dex in both
cases, and conclude that this scatter is within the uncertainties as
expected from Aller (\cite{aller84}). However, it has to be noted that
the intrinsic weakness of \oii\ 732.0, 733.0~nm lines, with respect to
the
\oii\ 372.7, 372.9~nm, produces larger errors in the abundance determination.

We adopted the prescriptions by Izotov et al.~(\cite{izotov94}) to
derive the helium ionic abundances, and by Izotov et
al.~(\cite{izotov05}) to derive the oxygen, nitrogen, sulfur, and
argon ionic abundances and ICFs.  Uncertainties on the ionic and total
abundance were computed with a standard propagation of the errors on
the flux measurements, which propagates to the de-reddening procedure,
to the electron temperature and density derivations, and to the ICFs
computation.  The ionic and total abundances with their errors are
reported in Table~\ref{Tab_chem}.

\section{The metallicity gradients in M~33}
\label{sect_grad}
\subsection{Helium}
In Fig.~\ref{Fig_he} we present the determinations of the ionic
He$^{+}$/H$^{+}$ abundance as a function of galactocentric distance.
There is no clear indication of radial gradient. While the range of
the He$^{+}$/H$^{+}$ abundance found here appears consistent with
previous works, the correction for neutral helium is very uncertain
and therefore the helium abundance gradient is difficult to derive.
Nonetheless, assuming the ICF scheme of He based on the $\eta$
parameter (e.g. Pagel et al.~\cite{pagel92}; V\'{\i}lchez et
al.~\cite{vilchez88}), we can put a limit for those \hii\ regions at
radii lower than 3~kpc for which a value of $\eta$ can be derived; for
these HII regions an ionization correction factor, ICF(He)$>$ 1.2 is
suggested by our observations. Further observations are needed to
confirm these claims.

\begin{figure}
\resizebox{\hsize}{!}{\includegraphics[angle=-90]{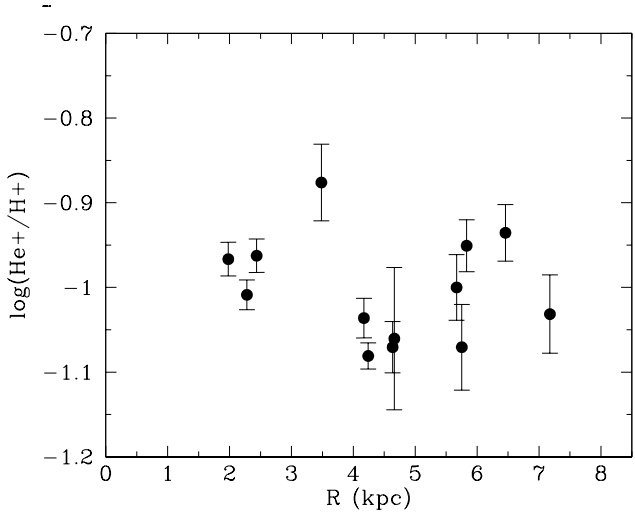}}
\caption{The He+/H+ abundance versus  galactocentric distance.}
\label{Fig_he}
\end{figure}

\subsection{Oxygen}
\begin{figure}
\resizebox{\hsize}{!}{\includegraphics[angle=-90]{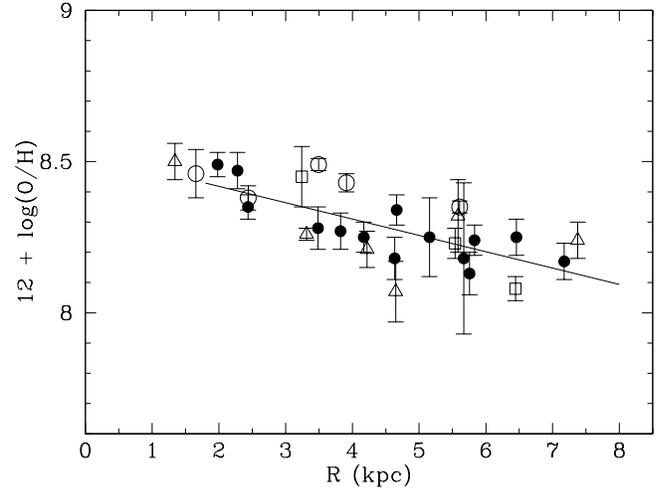}}
\caption{The O/H abundance (with electron temperature derived from observations)
versus the galactocentric distance:
{\em filled circles} from the present work, {\em empty squares} from Kwitter \&
Aller~(\cite{kwitter81}), {\em empty circles} from V\'{\i}lchez et
al.~(\cite{vilchez88}), and {\em triangles} from Crockett et
al.~(\cite{crockett06}).  The solid line is the weighted linear
least-squares fit to our 14 \hii\ regions.}
\label{Fig_oxy}
\end{figure}

In Fig.~\ref{Fig_oxy}, we show the oxygen abundance as a function of
galactocentric distance for  the 14 \hii\ where we measured the chemical
abundances.  A weighted linear least-square fit to our data (solid
line) gives a gradient:
\begin{equation}
12 + \log {\rm O/H} = -0.054 (\pm 0.011) \times  R + 8.53 (\pm 0.05).
\end{equation}
where R is the deprojected galactocentric distance in kpc, computed
assuming a distance of 840~kpc, an inclination of 53\arcdeg, and a
parallactic angle of 22\arcdeg.

The figure also shows the oxygen abundances, re-derived using the
Izotov's et al.~(\cite{izotov05}) ionic abundance and ICF formulae,
from all previous spectroscopic observations which allowed a direct
determination of the electronic temperature (Kwitter \& Aller
\cite{kwitter81}, V\'{\i}lchez et al.~\cite{vilchez88}, and Crockett
et al.~\cite{crockett06}, as shown in Table~\ref{Tab_chem_prev}).  In
case of multiple measurements of the same \hii\ region we have adopted
spectroscopic data from the most recent observations.  Where both
\oii\ 372.7 and 732.5 emission lines were available from the
literature (four \hii\ regions), we recomputed the ionic O$^+$ with both set of lines. 
The agreement is very
good, as shown in Table~\ref{Tab_chem_prev}.  Considering the whole
sample of oxygen measurements, the gradient still has a slope of $\sim
-0.06$ dex~kpc$^{-1}$, consistent with our determination and with
gradients measured in other non-barred galaxies of similar
morphological type (e.g. NGC~2403, Garnett et al.~\cite{garnett97}).

\subsection{Nitrogen}
In Fig.~\ref{Fig_nit} the radial variation of the nitrogen abundance is shown.
A weighted linear fit  of our data gives a gradient
\begin{equation}
12 + \log {\rm N/H} = -0.06 (\pm 0.02) \times R + 7.3 (\pm 0.1)
\end{equation}
similar to the oxygen one.  If also previous determinations are
considered, the overall slope of the gradient, results to be a bit steeper, namely $\sim -0.10$~dex~kpc$^{-1}$.

\begin{figure}
\resizebox{\hsize}{!}{\includegraphics[angle=-90]{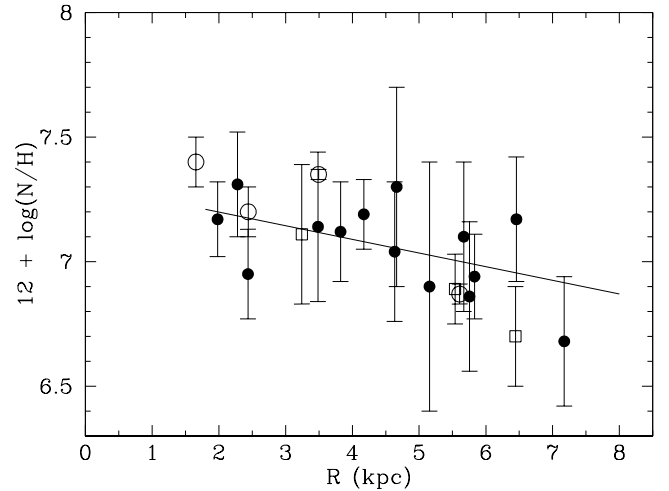}}
\caption{The 12+$\log$(N/H) abundance vs. galactocentric distance
(symbols and line types as in Fig.~\ref{Fig_oxy}).}
\label{Fig_nit}
\end{figure}

In Fig~\ref{Fig_no}, we present the relationship of N/O and the oxygen
abundance.  It appears that in M33 the N/O ratio is constant within
errors, with a value around $\log$(N/O)= -1.2, in between solar (-0.9,
Asplund et al.~\cite{asplund05}) and the value for dwarf irregular
galaxies ($\log$(N/O)= -1.5).  This probes that N/O along galactic
disks does not always follow the relation N/O proportional to O/H, as
expected for a purely secondary enrichment behaviour.  The almost
constant value of N/O with respect to O/H in M33 warns us against the
use of abundance determination which are based on models assuming that
the N/O ratio is proportional to O/H, and then using \nii/\oii\, to
infer the metallicity.  M33 shows that at least in some galaxy disks
this is not true.

\begin{figure}
\resizebox{\hsize}{!}{\includegraphics[angle=-90]{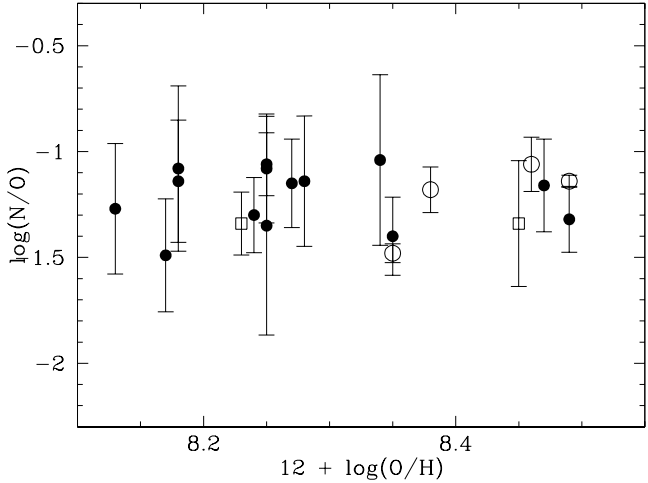}}
\caption{$\log$N/O vs. 12+$\log$(O/H) (symbols and line types as in Fig.~\ref{Fig_oxy}). }
\label{Fig_no}
\end{figure}

\subsection{Sulfur}
\begin{figure}
\resizebox{\hsize}{!}{\includegraphics[angle=-90]{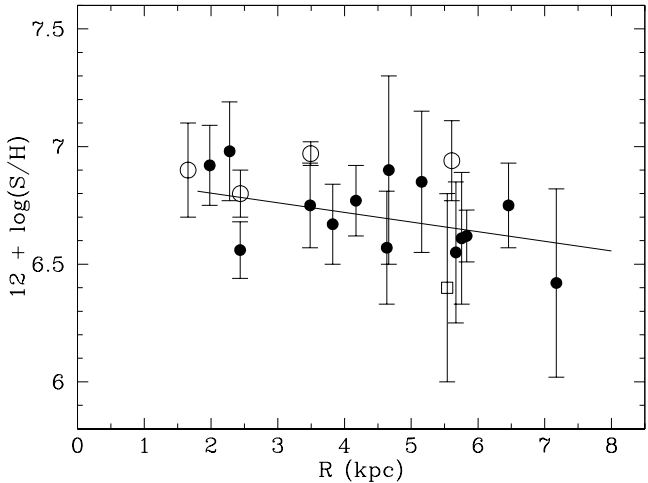}}
\caption{The 12+$\log$(S/H) abundance vs. galactocentric distance (symbols and line types as in Fig.~\ref{Fig_oxy}).}
\label{Fig_sul}
\end{figure}

Our spectral range did not allow the observation of the infrared
\siii\ emission lines, but the auroral line \siii\ 631.2~nm and \sii\ 671.7, 673.0~nm were
observed in a large number of \hii\ regions of our sample.  This
allowed us to measure the total sulfur abundance, since the larger
amount of sulfur is in the form of S$^{+}$ and S$^{2+}$, while the other ions
S$^{3+}$ and S$^{4+}$ contribute less.  The total S/H was
however computed taking also into account of the correction for the
unseen ionization stages, mainly S$^{3+}$,  using the ICF(S) by Izotov et al.~(\cite{izotov05}).
On the other hand, errors
on the total sulfur abundance measured via the \siii\ 631.2~nm and \sii\ 671.7, 673.0~nm  lines are larger
than via IR lines, since these emission lines are generally weaker.
Typical errors, in the range 0.1-0.4~dex, are  indicated in
Table~\ref{Tab_chem}.  The gradient of sulfur, as expected, resembles
that of oxygen, since these elements are both produced in stars
belonging to the same mass range (M$>$10~M$_\odot$).  A weighted
linear fit of our data gives

\begin{equation}
12 + \log {\rm S/H} = -0.04 (\pm 0.02) \times R + 6.88 (\pm 0.11).
\end{equation}

\section{Discussion}
\label{sect_discu}

The wealth of information collected during the last twenty years on
the metallicity distribution in the interstellar medium and young
stars of M33 allowed us to draw a general picture of its present time
gradient.

Individual studies, with small samples of objects, would lead to
significantly different - and thus uncertain - conclusions.  In fact,
in contrast with the result by Garnett et al.~\cite{garnett97},
indicating a mean gradient through the whole disk of M33 of
$-0.11$~dex~kpc$^{-1}$, more recent studies suggest a shallower
metallicity gradient derived for Ne and O.  Since both elements are
produced mainly by short-lived, massive stars ($M > 10$~$M_\odot$),
the slope of their gradient should be very similar.  In our Galaxy and
in nearby galaxies, the abundances of these two elements are generally
closely correlated, as supported by abundance measurements in Galactic
and extragalactic PNe (Henry~\cite{henry90}), showing that Ne/O is in
a good approximation constant over a wide range of O/H values.  This
constancy of oxygen and neon is, however, in apparent contrast with
the results obtained by Willner \& Nelson-Patel~(\cite{willner02}),
who derived Ne abundances for 25
\hii\  regions in M33 from infrared spectroscopy.
The best fit to the complete sample of their data is a step function
in a diagram showing the abundance of neon (relative to the solar one)
versus the galactocentric distance: $-0.15$~dex from 0.7 to 4~kpc, and
$-0.35$ from 4 to 6.7~kpc.  Avoiding the two outermost \hii\ regions
of their sample they found a best-fit slope
-0.05$\pm$0.02~dex~kpc$^{-1}$.  An even shallower slope for the Ne/H
gradient was found by Crockett et al.~(\cite{crockett06}),
$-0.016$~dex~kpc$^{-1}$.

In order to avoid peculiarity effects due to the limited size of
different samples, we have considered, in addition to the present work
results, oxygen measurements in a large number of objects
representative of the present time ISM metallicity: \hii\ regions
(optical spectroscopy\footnote{for these \hii\ regions we have
re-computed uniformly chemical abundances using the Izotov's et
al.~(\cite{izotov05}) ionic abundance and ICF formulae, only for the
\hii\ regions where the electron temperature diagnostic lines were
available (see Table~\ref{Tab_chem_prev})} by Kwitter \&
Aller~\cite{kwitter81}, V\'{\i}lchez et al.~\cite{vilchez88}, Crockett
et al.~\cite{crockett06}, and infrared spectroscopy\footnote{Neon
abundance were converted into oxygen using the solar ratio O/Ne by
Asplund et al.~(\cite{asplund05})} by Willner \&
Nelson-Patel~\cite{willner02}) and young stars (Monteverde et
al.~\cite{monteverde97}, Urbaneja et al.~\cite{urbaneja05}, Beaulieu
et al.~\cite{beaulieu06}).  The abundances were averaged in bins 1~kpc
wide, starting from the center of the galaxy.  The errors are the {\em
rms} scatter of the average value in each bin.

\begin{figure}
\resizebox{\hsize}{!}{\includegraphics[angle=-90]{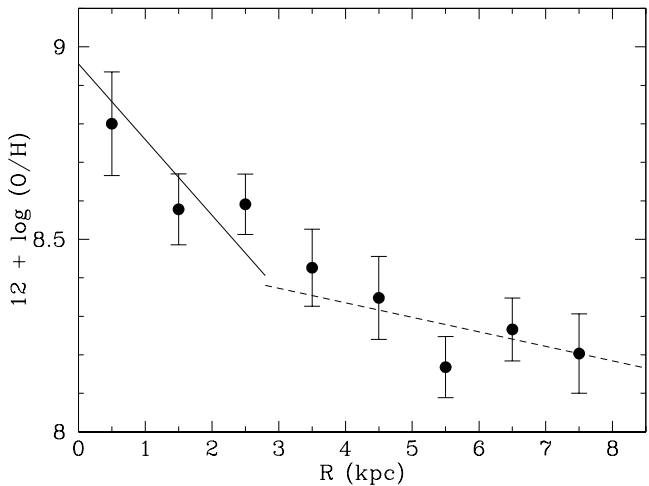}}
\caption{Mean oxygen abundance from \hii\ regions and young stars as indicated in the text
vs. galactocentric distance. The best-fit is a linear fit with two
slopes: -0.19$\pm$0.08~dex~kpc$^{-1}$ for R$<$3~kpc ({\em solid line})
and -0.038$\pm$0.015~dex~kpc$^{-1}$ ({\em dashed line}) for
R$\geq$3~kpc. }
\label{Fig_hist}
\end{figure}

From Fig.~\ref{Fig_hist}, it is clear the oxygen gradient cannot be
fitted with a single slope, that would exclude the inner points with
higher metallicity.  Two different slopes in the central and outer
regions are needed instead.  The best fit gives for the region from
the center to 3~kpc
\begin{equation}
 12 + \log {\rm O/H} = -0.19 (\pm 0.08) \times R + 8.95 (\pm 0.13)
\end{equation}
and for the radial region from 3 to 8 kpc
\begin{equation}
 12 + \log {\rm O/H} = -0.038 (\pm 0.015) \times R + 8.49 (\pm 0.08).
\end{equation}

The existing conflict on the determination of the slope of the M~33
metallicity gradient might be solved using the result of
Fig.~\ref{Fig_hist}, which involves all the best chemical abundance
determinations in this galaxy, including nebular abundances with
direct T$_e$ measurement and stellar abundances.  Stasinska et
al.~(\cite{stasinska06}) discussed the reliability of chemical
abundance derived in high metallicity \hii\ regions: even if the
\oiii\ 436.3 line is measured.  the derived O/H might be anyhow
underestimated since strong temperature gradients inside the nebula
due to the cooling by \oiii\ 52, 88~$\mu$m make the
\oiii\ 436.3/500.7 ratio overestimate the temperature.
Nonetheless this effect discussed
is expected to be significant only for abundances larger
than the solar value, in the case of M~33 this refers possibly
to the central kpc.
Thus the presence of stellar abundances, which are not affected by the same problems of
nebular abundances in the high metallicity regime,
makes the result of  Fig.~\ref{Fig_hist} more robust.

Since the gradient in M~33 cannot be represented by a single linear
fit, studies undertaken in different radial regions would produce
extremely different results: if inner regions were included, a steeper
gradient would result. On the contrary if only regions far form the
center were considered, a flatter gradient would be obtained.  We
remind that the steepening of the inner gradient is also sustained by
individual studies (e.g. V\'{\i}lchez et al.~\cite{vilchez88},
Beaulieu et al.~\cite{beaulieu06}).  Moreover, the relatively shallow
slope of the outer metallicity gradient is confirmed by several recent
works, as the previously quoted papers by Willner \& Nelson-Patel
(\cite{willner02}), Urbaneja et al.~(\cite{urbaneja05}), Crockett et
al.~(\cite{crockett06}).  Also recent works are obtaining noteworthy
results, as the work by Viironen et al.~(\cite{viironen07}) and that
by Rosolowsky \& Simon (\cite{rosolowsky07}).  The former is studying
via photoionization modelling the dependence on metallicity of
location of \hii\, regions in the diagnostic diagram \nii/\ha\, vs
\sii/\ha. For M33, using the sample of \hii\ regions presented in this
work and data of literature, they found a gradient across all the disk
of -0.05$\pm$0.01~dex~kpc$^{-1}$.  The latter work, called ``M33
Metallicity Project'', is studying a large number of
\hii\ regions in a South-West region of M33 via optical spectroscopy.
They measure an oxygen abundance gradient of -0.03~dex~kpc$^{-1}$.

From a theoretical point of view, Magrini et al.~(\cite{magrini07})
present in an accompanying paper a detailed model of the chemical
evolution of the disk of M 33. The model is able to reproduce the
observed gradients, together with other observables, as the gas and
star distribution, and the star formation rate. The predicted oxygen
abundance gradient is -0.067~dex~kpc$^{-1}$, across the whole disk
excluding the inner 1~kpc regions\footnote{The model does not take
into account the central bulge and thus cannot reproduce the central
enhancement of the metallicity.}  in excellent agreement with the best
fit to the data in Fig.~\ref{Fig_hist} in the same radial region,
-0.07$\pm$0.01~dex~kpc$^{-1}$.

The relatively shallow slope and the apparent flattening with time of
the gradient obtained from the comparison of the present time gradient
with PN (Magrini et al.~\cite{magrini04}) and RGB stars (Barker et
al.~\cite{barker06}) gradients, are explained by the model with a slow
continuous formation of the disk by infall.  A conspicuous cloud of
neutral gas infalling into the disk of M33 recently detected at 21-cm
by Westmeier et al.~(\cite{westmeier05}) and the presence of carbon
stars found at larger radii where the drop of the \ha\ flux occurs
(Davidge \cite{davidge03}, Block et al. \cite{block04},
\cite{block06}, Rowe et al. \cite{rowe05}) support this picture,
suggesting that accretion of gas in the disk of M~33 is still taking place.

The results of this model and the new detailed observations available
for M~33 open new scenarios on the formation of spiral galaxies.
These scenarios are not in opposition to the ``classical'' models of
galactic disk formation, i.e. rapid collapse of an initial halo, but
they extend them with the possibility of further frequent accretion
events from mergers and interactions in the intergalactic medium.  A
full discussion of chemical evolution modeling for M~33 is given by
Magrini et al.~(\cite{magrini07}).

\section{Conclusions}
\label{sect_conclu}
We analyzed optical spectra of 72 emission-line objects in M~33,
including mostly \hii\ regions, but also five SNRs and two candidate
PNe.  Most of the emission-line objects are in the low density limit,
with n$_e <$100 cm$^{-3}$, only three of them have higher density.
The direct T$_e$ determination was possible for 15 \hii\ regions by
the measurements of the \oiii\ 436.3~nm line and in three of them of
the \nii\ 575.5~nm line.

The radial variation of several emission-line ratios does not appear
to be much affected by the different morphology of the \hii\ regions,
even if in the \oiii/\hb\, diagram the higher surface brightness
sources show a higher correlation, signature of spectra dominated by
local chemical abundance effects.  Emission-line ratio diagrams
allowed us to separate the different types of emission-line objects
and to infer the presence of metallicity gradients.

For the 14 \hii\ regions with determined T$_e$ and with detection of
both \oii\ and \oiii\ lines we derived chemical abundances using the
ICF methods.  We found metallicity gradients in good agreement with
previous determinations in the same radial regions. The resulting
global oxygen abundance gradient is -0.06$\pm$0.01~dex~kpc$^{-1}$.

We collected oxygen measurements in a large number of objects
representative of the present time ISM metallicity to draw a complete
picture of the shape and magnitude of the metallicity gradient.  We
found the best fit to the data using two different slopes,
corresponding to the central regions R$<$3~kpc
(-0.19$\pm$0.08~dex~kpc$^{-1}$) and outer regions R$\geq$3~kpc
(-0.038$\pm$0.015~dex~kpc$^{-1}$).  The shallower outer gradient is in
agreement with several new works and it can be explained with a slow
continuous accretion of the disk of M~33 from the intergalactic
medium.

\begin{acknowledgements}
R.L.M.C., A.M., and P.L. acknowledge financial support from the
Spanish Ministry of Science and Education (grant AYA02-00883).  The
work of L.M. is supported by a INAF post-doctoral grant 2005.  We are
grateful to Edvige Corbelli for many useful discussions about M33, and
for suggestions about the radial behaviour of the extinction.
\end{acknowledgements}

\scriptsize{
\longtab{1}{
\begin{longtable}{rllllll}
\caption{The list of emission line objects spectroscopically observed in M~33,
including \hii\ regions, SNRs and PNe. The identification names are
from: BCLMP-- Boulesteix et al. (\cite{boulesteix74});
CPSDP-- Courtes et al. (\cite{courtes87});
VGVC-- Viallefond et al. (\cite{viallefond86});
GDK99-- Gordon et al. (\cite{gordon99});
MCM00- Magrini et al. (\cite{magrini00}); CDL04-- Ciardullo et
al. (\cite{ciardullo04}).  The emission-line objects not identified in
previous works, mostly small isolated \hii\ regions, are labeled with
LGC-HII-n, standing for \hii\ regions discovered by the Local Group
Census project (cf. Corradi \& Magrini \cite{corradi06}).  The
coordinates J2000.0 of the position of the fibers projected on the sky are
shown in the second and third columns. They do not correspond exactly
to the center of the emission line objects, but generally to the
maximum \oiii\ emissivity.  The observed photometric fluxes  in unit of
10$^{-15}$erg/cm$^2$, F$_{H\alpha}$,  and the area in arcsec$^2$
were taken from Hodge et al.~(\cite{hodge02}), when this information is available,
otherwise their were measured in the LGC INT images (columns 4$^{th}$, 5$^{th}$).
The last column shows
the main characteristic of the morphology: {\em e} extended, {\em c}
small/isolated, {\em k} high surface brightness knot in an extended complex, {\em s}
shell/ring-like.  }\label{Tab_pos_hii}\\
\hline\hline
Id & Name     & RA        & Dec             & F$_{H\alpha}$         & Area & Morphology\\
   &          & \multicolumn{2}{c}{J2000.0} & (10$^{-15}$erg/cm$^2$s) & (arcsec$^2$)           &           \\
\hline
\endfirsthead
\caption{continued.}\\
\hline\hline
Id & Name     & RA        & Dec             & F$_{H\alpha}$         & Area & Morphology\\
   &          & \multicolumn{2}{c}{J2000.0} & (10$^{-15}$erg/cm$^2$s) & (arcsec$^2$)          &           \\
\hline
\endhead
\hline
\endfoot
3  &BCLMP~275a & 1:32:29.5 & 30:36:08        & 80                 &  187          & e   \\
4  &LGC-HII-1  & 1:32:30.2 & 30:32:35	      & 15                 &  150          & c     \\
7  &BCLMP~272  & 1:32:33.4 & 30:27:45        & 520                &  606          & e  \\
8  &BCLMP~273b & 1:32:33.5 & 30:30:43        & 77                 &  207          & e  \\
14 &BCLMP~266  & 1:32:40.8 & 30:24:25        & 270                &  497          & k     \\
17 &BCLMP~263a & 1:32:41.8 & 30:21:17        & 20                 &  92           & k     \\
19 &LGC-HII-2  & 1:32:43.0 & 30:19:31        & 20                 &  13           & c     \\
22 &BCLMP~238  & 1:32:44.5 & 30:34:54        & 270                &  317          & e   \\
\end{longtable}
}

\scriptsize{
\longtab{2}{
\begin{longtable}{lllllllllll}
\caption{Observed line fluxes $F_\lambda$ of seventy-two emission-line objects, including mainly \hii\ regions (see Tab.\ref{Tab_pos_hii}).
$F_\lambda$ are represented on a scale where F(\hb)=100.0.
 `$<$' indicates upper limits. Associated errors are described in the
text.  Some nebular quantities are also reported: the extinction coefficient, \cbeta, the electron
density, n$_e$ and the electron temperatures, T$_e$(\oiii) and T$_e$(\nii)}\label{Tab_hii_flux}\\
\endfirsthead
\caption{Continued.}\\
\hline\hline
\hline
\endhead
\hline
\endfoot
\hline\hline
 		 & BCLMP275a     &LGC~HII~1    &BCLMP272     &BCLMP273a     &BCLMP266     &BCLMP263a&LGC~HII~2    &BCLMP238      & LGC~HII~3   & BCLMP240a    \\
\noalign{\smallskip}
\noalign{\smallskip}
\hline
\noalign{\smallskip}
\cbeta\		  &0.06$\pm$0.02 &0.40$\pm$0.08&0.30$\pm$0.02&0.40$\pm$0.04 &0.37$\pm$0.01 & 0.08$\pm$0.02 &0.00$\pm$0.01&0.00$\pm$0.01 &0.27$\pm$0.01 & 0.04$\pm$0.02	\\
T$_e$(\oiii)(K)  & $<$15000     &-            &$<$17000     &-             &$<$12000      & $<$14500      &10500$\pm$500 &$<$13000&11600$\pm$400& $<$11000  \\
n$_e$(cm$^{-3}$)  &  $<$10       &1800$\pm$1000& $<$10       &$<$10         & $<$10        & $<$30            &$<$10      &$<$10          & $<$10  &$<$10            \\
 \noalign{\smallskip}
\hline
\noalign{\smallskip}
410.1    \hd\     & 25$\pm$2	 & -	       & 20$\pm$3    & -	    &17$\pm$2 	   & 19$\pm$2      & 27$\pm$2    &29$\pm$2      &24$\pm$1     & 21$\pm$3   \\
434.0    \hg\     & 44$\pm$2 	 & -	       & 44$\pm$4    & -	    &37$\pm$ 	   & 40$\pm$3      & 48$\pm$2    &43$\pm$2      &51$\pm$1     & 45$\pm$2   \\
436.3    \oiii\   & $<$4  	 & -	       & $<$3	     & -	    &$<$4 	   & $<$2          & 1.3$\pm$0.3 &$<$1.6        &2.4$\pm$0.6  &$<$ 1.7    \\
447.1    \hei\    & -   	 & -	       & -	     & -	    &- 	           & -             & -           &-             &5$\pm$1      & -      \\
\end{longtable}
}
}

\longtab{3}{
\begin{longtable}{llllll}
\caption{ Total and ionic chemical abundances, and ionization correction factors  of fourteen \hii\ regions.
The galactocentric distance in kpc is also reported.} \label{Tab_chem}\\
\endfirsthead
\caption{Continued.}\\
\hline\hline
\hline
\endhead
\hline
\endfoot
\hline\hline
					& LGC~HII~2 	          & LGC~HII~3   & BCLMP289      & BCLMP218       & MCM00Em24                  \\
radial distance (kpc)				& 6.5                     & 5.8         & 5.2           & 4.2            & 5.7                          \\
\noalign{\smallskip}
\hline
\noalign{\smallskip}
He$^+$/H$^+$            		&0.116$\pm$0.009   	  &0.112$\pm$0.008&0.12$\pm$0.04&0.092$\pm$0.005 &0.100$\pm$0.009   		\\
\hline

\noalign{\smallskip}
N$^+$/H$^+$$\times$10$^{5}$		&0.86$\pm$ 0.06	  	  &0.50$\pm$0.03&0.41$\pm$0.07 &0.65$\pm$0.02	 &0.66$\pm$ 0.06   	\\
ICF(N)					&1.73	  	  	  &1.76 	&1.82		&2.39		 &1.9             \\
\noalign{\smallskip} 28
N/H$\times$10$^{5}$			&1.5$\pm$0.9	  	  &0.88$\pm$0.34&0.8$\pm$0.9	&1.5$\pm$0.5	 &1.3$\pm$0.5     	\\
12 + $\log$ N/H			        &7.17$\pm$0.25     	  &6.94$\pm$0.17&6.90$\pm$0.50	&7.19$\pm$0.14	 &7.10$\pm$0.30   	\\

\hline

\noalign{\smallskip}
O$^+$/H$^+$$\times$10$^{5}$		&12$\pm$5	          & 12$\pm$3   &11.8$\pm$10	&8.9$\pm$2.9	 &9.5$\pm$10     	\\
\end{longtable}
}

\longtab{4}{
\begin{longtable}{llllll}
\caption{ Total and ionic chemical abundances, and ionization correction factors  of \hii\ regions
with line fluxes from previous observations and with
electron temperature diagnostic lines detected: KW81 (Kwitter \& Aller \cite{kwitter81}), V88 (Vilchez et al.~\cite{vilchez88}),
C06 (Crockett et al.~\cite{crockett06}).
In case of multiple observations by different authors we used the most recent ones.
The galactocentric distance in kpc is also reported.} \label{Tab_chem_prev}\\
\endfirsthead
\caption{Continued.}\\
\hline\hline
\hline
\endhead
\hline
\endfoot
\hline\hline
				  & NGC595 (V88)   & MA2 (V88)     & NGC604 (V88)   & IC131 (V88)        & NGC588 (V88)                  \\
radial distance (kpc)		  & 1.7            & 2.4           &3.5             & 3.9                & 5.6                        \\
\noalign{\smallskip}
\hline
\noalign{\smallskip}
He$^+$/H$^+$            	  &0.07$\pm$0.05   &0.075$\pm$0.003&0.082$\pm$0.002 &0.080$\pm$0.009   	 &0.09$\pm$0.01   	\\
\hline

\noalign{\smallskip}
N$^+$/H$^+$$\times$10$^{5}$	  &1.43$\pm$0.08&0.70$\pm$0.05     &0.93$\pm$0.02	 &0.64$\pm$ 0.07& 0.18$\pm$0.02      	\\
ICF(N)				  &2.0  	&2.3		   &2.4 		 &3.1           & 4.0               \\
\noalign{\smallskip}
N/H$\times$10$^{5}$		  &2.9$\pm$0.6  &1.6$\pm$0.4	   &2.2$\pm$0.1	         &2.0$\pm$0.4   & 0.74$\pm$0.65     	\\
12 + $\log$ N/H			  &7.4$\pm$0.1  &7.21$\pm$0.10	   &7.35$\pm$0.02	 &7.30$\pm$0.09 & 6.87$\pm$0.04     	\\

\hline

\noalign{\smallskip}
O$^+$/H$^+$$\times$10$^{5}$ (372.7)& 17$\pm$2.5   &12.5$\pm$2	   &15.4$\pm$0.5	 &9.9$\pm$1.2   & 5.8$\pm$0.3       	\\
\end{longtable}
}

\end{document}